\documentclass[twocolumn,times,tighten]{aastex62}
\usepackage{amsmath,amstext}

\usepackage{newtxmath} 

\usepackage{epsfig}
\usepackage{float}
\usepackage{natbib}
\usepackage[figure,figure*]{hypcap}

\def\Snospace~{\S{}}

\def\chandra{\emph{Chandra}}
\def\vla{\emph{VLA}}
\def\as{$^{\prime\prime}$}
\def\am{$^{\prime}$}

\defcitealias{2006ESASP.604..723M}{M06}
\defcitealias{2007PhR...443....1M}{MV}
\defcitealias{2012MNRAS.423..236R}{R12}
\defcitealias{2016ApJ...833...99W}{W16}
\defcitealias{2005ApJ...627..733M}{M05}

\shorttitle{Shock front and radio halo in Abell 520}
\shortauthors{Wang, Giacintucci, \& Markevitch}

\begin{document}

\title{BOW SHOCK IN MERGING CLUSTER A520: THE EDGE OF THE RADIO HALO AND THE \\
  ELECTRON--ION EQUILIBRATION TIMESCALE}

\author{Qian H. S. Wang}
\affiliation{Department of Astronomy, University of Maryland, College Park, MD 20742, USA}

\author{Simona Giacintucci}
\affiliation{Naval Research Laboratory, 4555 Overlook Avenue SW, Code 7213, Washington, DC 20375, USA}

\author{Maxim Markevitch}
\affiliation{Astrophysics Science Division, NASA Goddard Space Flight Center, Greenbelt, MD 20771, USA}
\affiliation{Joint Space-Science Institute, University of Maryland, College Park, MD, 20742, USA}

\begin{abstract}
We studied the prominent bow shock in the merging galaxy cluster A520 using a
deep \chandra\ X-ray observation and archival \vla\ radio data. This shock is
a useful diagnostic tool, owing to its clear geometry and relatively high Mach
number. At the ``nose'' of the shock, we measure a Mach number of
$M=2.4_{-0.2}^{+0.4}$. The shock becomes oblique away from the merger axis,
with the Mach number falling to $\simeq$1.6 around 30$^{\circ}$ from the
nose. The electron temperature immediately behind the shock nose is
consistent with that from the Rankine-Hugoniot adiabat, and is higher (at a
95\% confidence) than expected for adiabatic compression of electrons followed
by Coulomb electron--proton equilibration, indicating the presence of
equilibration mechanisms faster than Coulomb collisions. This is similar to an
earlier finding for the Bullet cluster. We also combined four archival \vla\
datasets to obtain a better image of the cluster's giant radio halo at 1.4 GHz. An
abrupt edge of the radio halo traces the shock front, and no emission is detected
in the pre-shock region. If the radio edge were due only to adiabatic compression
of relativistic electrons in pre-shock plasma, we would expect a pre-shock radio
emission detectable in this radio dataset; however, an interferometric artifact
dominates the uncertainty, so we cannot rule this model out. Other
interesting features of the radio halo include a peak at the remnant of the cool core,
suggesting that the core used to have a radio minihalo, and a peak marking a
possible region of high turbulence.
\end{abstract}


\section{Introduction}
Galaxy clusters form and grow via mergers of less massive systems. During a
merger, shocks and turbulence are generated in the intracluster medium (ICM).
They dissipate the kinetic energy of the subclusters (the fraction carried by
the ICM) and heat the ICM to the temperature that allows thermal pressure to
balance the gravity of the combined massive cluster. Observing the details of
this process allows us to gain insight into the complex physics of the
magnetized intracluster plasma.

An interesting possibility to observe the process of energy equilibration of
the plasma electrons and protons is afforded by shock fronts.  In a simple
picture, for shocks with low Mach numbers $M$\/ typical for cluster mergers,
the shock passage heats ions dissipatively, while electrons, whose thermal
velocity is much higher than that of the shock, are compressed adiabatically.
They then equilibrate via Coulomb collisions with protons. If this indeed is
how the electron temperature $T_e$ behaves in clusters, this would have
far-reaching consequences --- for example, total mass estimates at large
cluster radii, based on the hydrostatic assumption and the electron
temperature (e.g., \citealt{1988xrec.book.....S}), would be biased low because
of an underestimate of the average temperature in the low-density cluster
outskirts (e.g., \citealt{1996ApJ...456..437M};
\citealt{1999ApJ...520..514T}). This effect has astrophysical implications far
beyond galaxy clusters --- e.g., certain models of accretion disks rely on
this timescale \citep{1982Natur.295...17R}.

In the X-ray, we directly observe only the electron temperature $T_e$, but at
an intracluster shock, we can deduce the equilibrium plasma temperature from
the directly observable gas density jump, which gives the Mach number.
Luckily, the cluster Mach numbers are low enough for the density jump to be
far from its asymptotic value. We can also determine the gas flow velocities
on both sides of the shock. We are further lucky that the typical ICM
densities and temperatures are such that the product of the Coulomb
electron--proton equilibration timescale and the sound speed is of the order of tens
of kiloparsecs, which is resolvable by \chandra. This allows us to derive an electron
temperature profile across the shock and see if it follows the prediction for
collisional equilibration in the narrow zone downstream from the shock. This
test has first been applied to the Bullet Cluster
(\citealt{2006ESASP.604..723M}, hereafter \citetalias{2006ESASP.604..723M};
see also \citealt{2007PhR...443....1M}, hereafter
\citetalias{2007PhR...443....1M}), who obtained a tantalizing conclusion
(though only at a 95\% significance)
that the equilibration timescale is much shorter than
Coulomb. If seen systematically in other cluster shocks, this may suggest the
presence of a faster equilibration mechanism in the hot magnetized ICM. While
shock fronts are also observed in supernova remnants and even in situ in the
solar wind, the electron--proton equilibration timescale can be studied so
directly only in cluster shocks, because of the favorable combination of the
linear scales and the Mach numbers (e.g., \citetalias{2007PhR...443....1M}).

This test requires a simple, reasonably unambiguous shock geometry and a
high-statistics, high-resolution X-ray dataset in order to derive a 3D
temperature jump at the shock. After the Bullet cluster result,
\citet[hereafter \citetalias{2012MNRAS.423..236R}]{2012MNRAS.423..236R}
examined two other merger shocks that fit these requirements, those in A2146,
but their results were inconclusive because of large uncertainties and the low
Mach number of one of the shocks (the difference between shock heating and
adiabatic heating of electrons becomes practically undetectable for $M\lesssim
2$). A deep \chandra\ observation of A520, which we have analyzed in
\citet[hereafter \citetalias{2016ApJ...833...99W}]{2016ApJ...833...99W} for
everything else other than the shock front, presents another one of those rare
opportunities. We will describe this test in \autoref{sec:equilibration}.

Shocks and turbulence generated by the merger would not only heat the
intracluster gas, but also accelerate ultrarelativistic particles and amplify
magnetic fields that coexist with the thermal plasma. These ultrarelativistic
electrons reveal themselves through synchrotron radio emission in the shape of
radio halos and relics \citep[e.g.,][]{2005ApJ...627..733M, 2008A&A...486..347G,
2010Sci...330..347V, 2012A&ARv..20...54F,
2014IJMPD..2330007B}. While the energy density of these components is a small
fraction of the gas thermal pressure, they can substantially alter the physics
of the ICM. A520 exhibits a giant radio halo detected by Very Large Array
\citep[\vla;][]{2001A&A...376..803G, 2014A&A...561A..52V}, whose distinct
brightness edge coincides with the X-ray shock front \cite[hereafter
\citetalias{2005ApJ...627..733M}]{2005ApJ...627..733M}, similar to several
other clusters with shock fronts \citep[e.g.,][]{2008A&A...486..347G,
2012mgm..conf..397M, 2013A&A...554A.140P, 2014MNRAS.440.2901S}. The previous
analyses of the A520 radio halo used two subsets of \vla\ data separately,
which limited the sensitivity both because of the partial statistical accuracy
and the limited coverage of the Fourier space (the {\em uv}\/ plane) by the
antennas during a typical \vla\ observation, which may lead to lower
reconstructed image fidelity. To take full advantage of the existing radio
data, we combine all archival \vla\ observations in \autoref{sec:radio}. We
revisit the earlier finding of the coincidence of the radio halo edge and the
X-ray shock front. We use the improved radio sensitivity to put an
upper limit on the radio emission in the pre-shock region and test
one of the possible mechanisms for the origin of the radio edge
considered in \citetalias{2005ApJ...627..733M} --- adiabatic compression of
pre-existing relativistic electrons. There are other illuminating coincidences
between the radio and X-ray structure of the cluster that we discuss in
\autoref{sec:radio_features}.

We assume a flat cosmology with $H_0=70$ km s$^{-1}$ Mpc$^{-1}$ and
$\Omega_m=0.3$, in which 1\as~ is 3.34 kpc at the cluster's redshift of
$0.203$. Uncertainties are quoted at 90\% confidence unless otherwise stated.

\section{X-Ray Data Analysis}
\label{sec:xray}

The \chandra\ data reduction is described in \citetalias{2016ApJ...833...99W},
where we discussed all the A520 features seen in this dataset other than the
shock front. In summary, we use 447 ks of \chandra\ observations of A520
performed in 2007--2008 (ObsIDs 9424, 9425, 9426, 9430). Three earlier
observations (ObsIDs 528, 7703, 4125) were not used because they would not
meaningfully improve our results while adding complexity to the analysis.
Spectral analysis was performed using XSPEC (v12.9.0). Temperatures were
obtained by fitting an absorbed APEC model, while accounting for the
additional background component as determined in
\citetalias{2016ApJ...833...99W}. The redshift was fixed at $z=0.203$, while
metal abundance (relative to \citealt{1989GeCoA..53..197A}) and Galactic
absorption were fixed at the best-fit cluster average values of 0.21 and
$N_H=6.3\times10^{20}\text{ cm}^{-2}$, respectively, obtained from a fit to
the spectrum from an $r=3'$ circle centered on the X-ray centroid.

\section{Radio Data Analysis}
\label{sec:radio}

We reanalyzed the archival \vla\ data at 1.4 GHz from project AF349 (the data
used in \citealt{2001A&A...376..803G}) and projects AC776 and AC706 (the data
used in \citealt{2014A&A...561A..52V}), which observed A520 in C and D array
configurations. \autoref{tab:vla} gives technical details of these
observations.

\begin{table*}[t!]
\caption{Details of archival \vla\ observations of A520}
\begin{center}
\begin{tabular}{lcccccccccc}
\hline\noalign{\smallskip}
\hline\noalign{\smallskip}
Project & Configuration & Frequency & Bandwidth & Date & Time  & FWHM, PA   &   rms  & Primary & $S_{\rm calibrator}$ \\
        &               &  (MHz)    &    (MHz)  &      & (min) & ($^{\prime \prime} \times^{\prime \prime}$, $^{\circ}$)\phantom{00}
  & ($\mu$Jy b$^{-1}$) & Calibrator & (Jy) \\
\noalign{\smallskip}
\hline\noalign{\smallskip} {}
AF349 & C & 1364.9/1435.1       & 50/50 & 1998 Dec 8  & 129 & $15.4\times14.9$, 59    & 25 & 3C48  & 16.4/15.7 \\
AF349 & D & 1364.9/1664.9$^{a}$ & 50/25 & 1999 Mar 19 & 180 & $50.6\times49.4$, 27    & 65 & 3C48  & 16.4/14.1 \\
AC776 & C & 1364.9/1435.1       & 50/50 & 2005 Aug 30 & 250 & $15.4\times14.5$, $-29$ & 22 & 3C147 & 23.1/22.2 \\
AC706 & D & 1364.9/1435.1       & 50/50 & 2004 Aug 13 & 345 & $48.8\times46.0$, 0     & 50 & 3C296 & 15.4/15.0 \\
\hline{\smallskip}
\end{tabular}
\end{center}
\label{tab:vla}
Notes. -- Column (1): {\em VLA}\/ project identifier; column (2): array
configuration; columns (3) and (4): frequency and width of the two intermediate
frequency (IF) channels used during the observation; columns (5) and (6):
observation date and total time on source; column (7): full width at
half-maximum (FWHM) and position angle (PA) of the beam; 
column (8): image rms noise; columns (9) and (10): primary calibrators and their
flux densities set according to the \citet{2013ApJS..204...19P} scale.\\
$^{\text{a}}$ We used only the 1364.9 MHz IF channel here.\\
\end{table*}

We calibrated and reduced the datasets separately using the Astronomical Image
Processing System (AIPS). We followed the standard procedure, with amplitude
and phase calibration carried out after accurate editing of the raw data on
both the primary and secondary calibration sources. The flux density scale was
set using the amplitude calibrators listed in \autoref{tab:vla} and the
\cite{2013ApJS..204...19P} coefficients in AIPS {\sc SETJY} task. The accuracy
of the flux density scale is estimated to be within 3\%. Phase-only
self-calibration was applied to each dataset to reduce the effects of residual
phase errors. Final images were made using the multi-scale {\sc CLEAN}
algorithm implemented in AIPS {\sc IMAGR} task, which results in better
imaging of the extended sources compared to the traditional single-resolution
{\sc CLEAN} \citep[e.g.,][]{2006AJ....131.2900C}. After self-calibration, we
combined the C and D data into a single data set. For the AF349
D-configuration observation, we used only the 1364.9 MHz IF channel that
matches the frequency and width of the first IFs of all other data sets.
Finally, a further cycle of phase calibration was applied to the combined
dataset to improve the image quality. We reached an rms sensitivity level of
20 $\mu$Jy beam$^{-1}$ in the final combined image, with a restoring beam of
$19\arcsec$.

\begin{figure}[!htb]
\vspace{3mm}
        \centering
        \leavevmode
        \includegraphics[width=80mm]
                {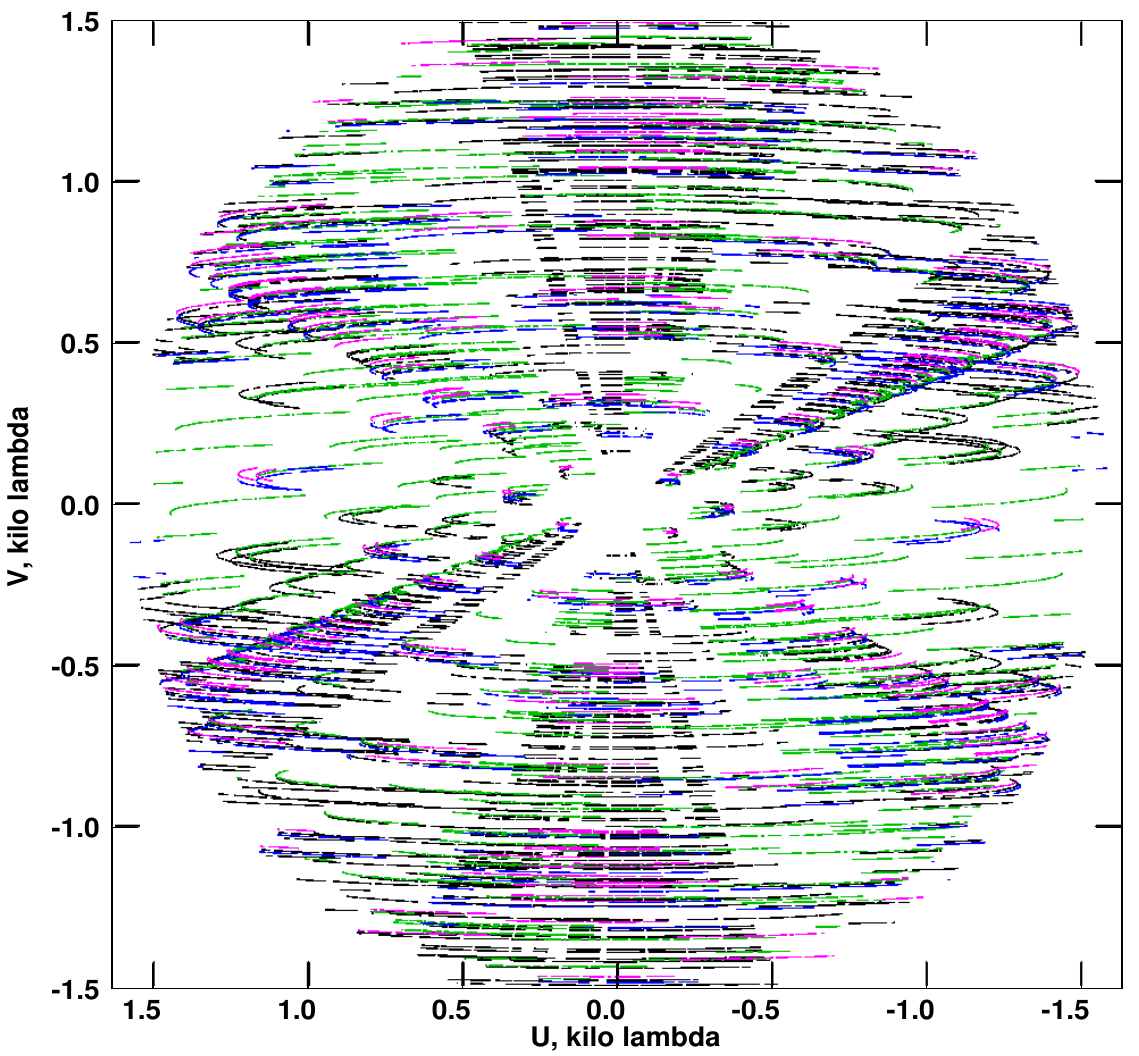}
\caption{Coverage of the $uv$\/ plane of spatial frequencies by the four combined
{\em VLA}\/ datasets (shown by different colors). Fuller coverage results in a
reconstructed image with higher fidelity for the extended features. The
datasets are complementary, especially at smaller wavenumbers near the center
of the plot (corresponding to larger angular scales).}
\label{fig:uv}
\end{figure}

Good sampling of short baselines (i.e., close antenna pairs) in the {\em uv}\/
plane is crucial for correct determination of the flux density, size, and
structure of a low-surface-brightness source like the radio halo in A520. The
inner portion of the {\em uv}\/ plane of our final combined data set is shown
in \autoref{fig:uv}. Only visibilities corresponding to baselines shorter
than 1.5 k$\lambda$ are plotted. The very good sampling of short spacings in
this plot ensures high-fidelity imaging of the radio halo whose angular size
of about 5$^{\prime}$ (diameter) is sampled by visibilities shorter than 0.8
k$\lambda$. Nominally, the largest structure detectable by this dataset should be
about 3 Mpc; we will return to this in a more quantitative way below.

\begin{figure*}[!ht]
        \centering
        \leavevmode
        \includegraphics[width=85mm,keepaspectratio,bb=50 160 535 640,clip]
                {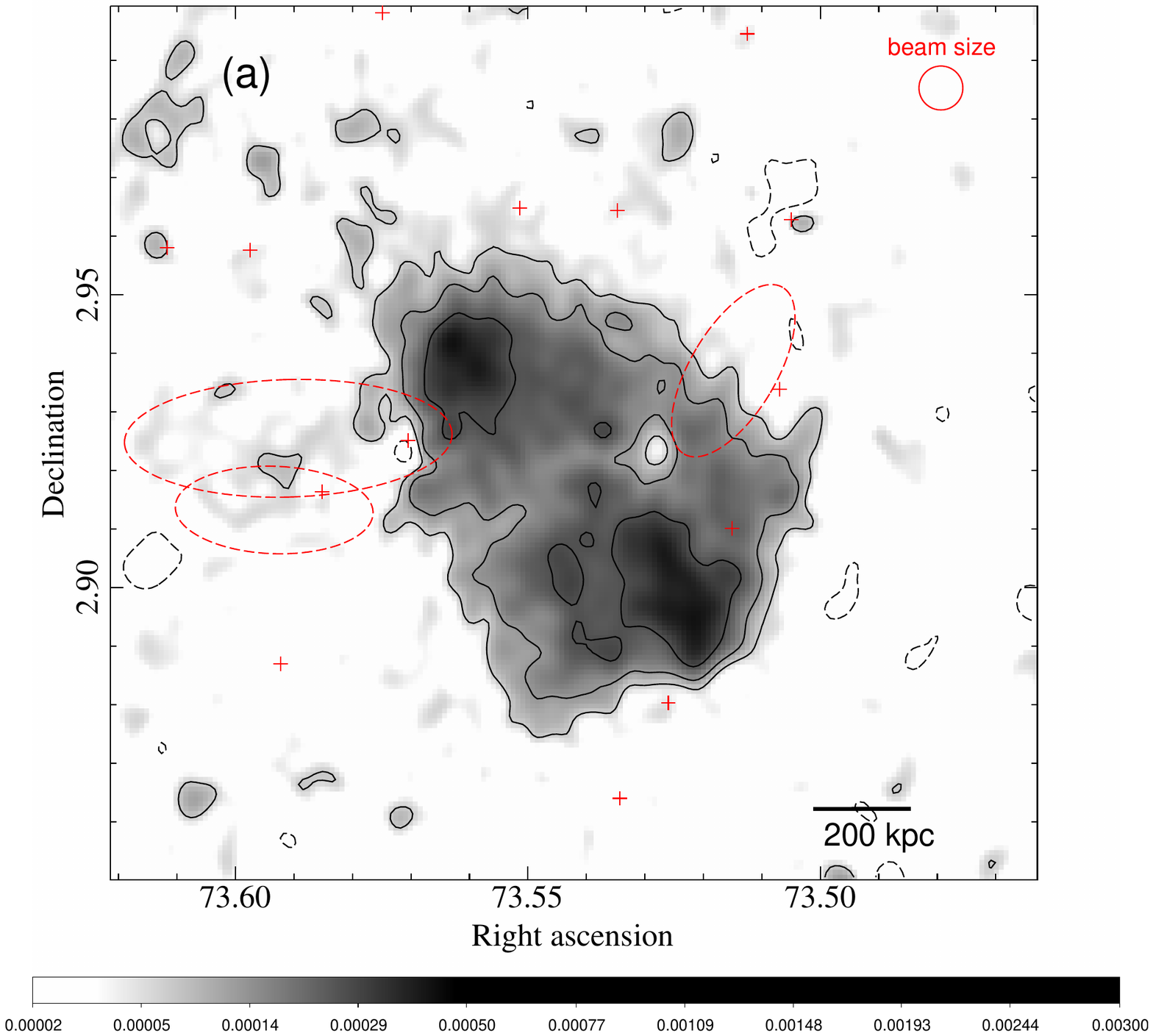}
        \hfil
        \includegraphics[width=85mm,keepaspectratio,bb=50 160 535 640,clip]
                {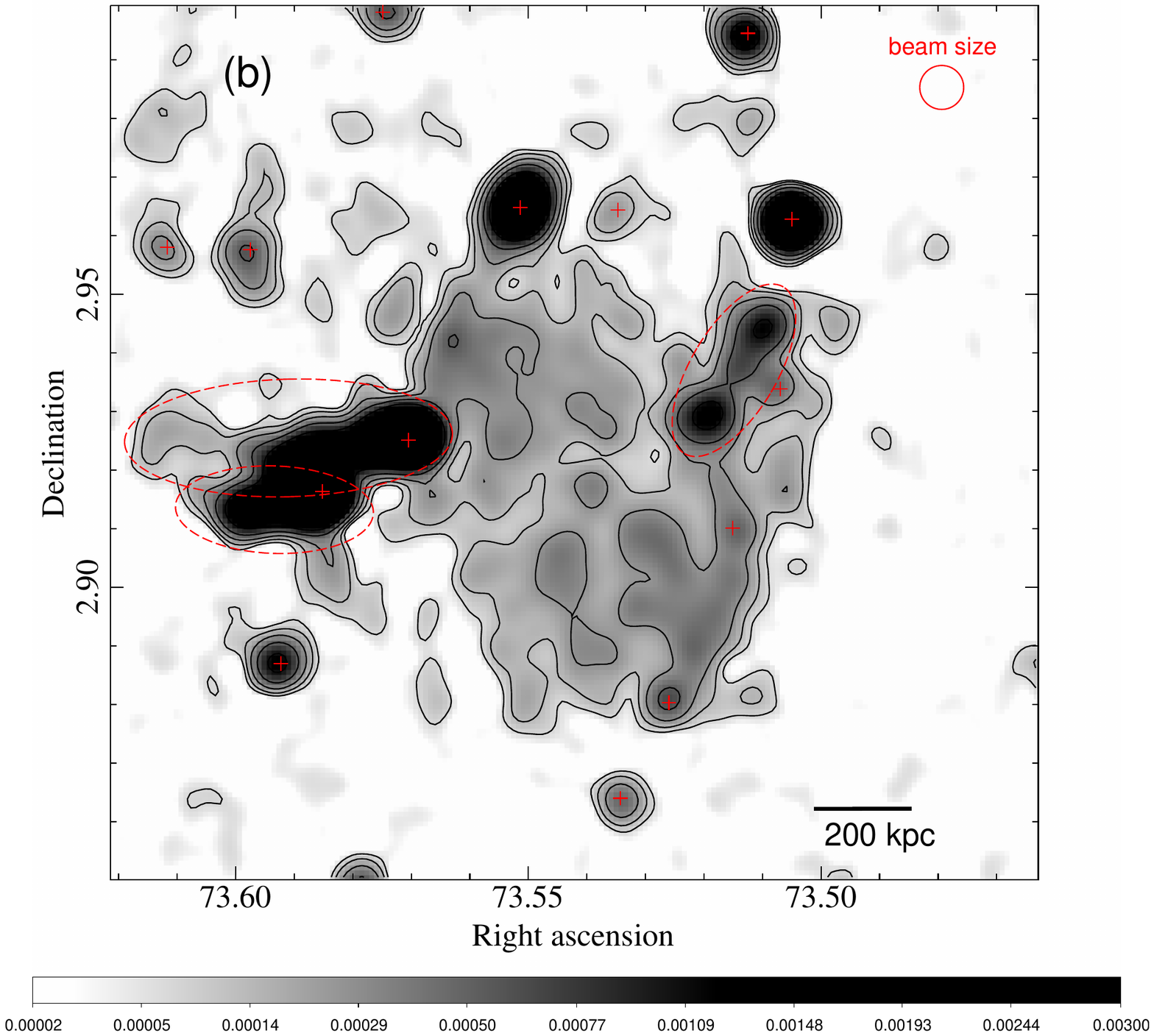}

        \caption{ ({\em a}) 1.4 GHz image of the radio halo after the removal
          of unrelated sources. Red crosses mark the positions of point
          sources while dashed ovals mark the three extended sources
          associated with radio galaxies.
          ({\em b}) Radio brightness image before the source removal.
          Radio contours start at 66~$\mu$Jy~beam$^{-1}$ ($3\sigma$) in steps
          of $\times2$ (dashed contours are negative). The beam size is the
          same in both images.
          200~kpc is 1\am.}
        \label{fig:radio}
\end{figure*}

To image only the extended and compact radio sources unrelated to the giant
halo, we produced images using only the baselines longer than 0.5~k$\lambda$
and longer than 1~k$\lambda$, respectively. We identified 16 such sources with
peak flux densities exceeding a $3\sigma$ level of 0.06~mJy~beam$^{-1}$ (for a
$19\arcsec$ restoring beam) within $r\sim 1$~Mpc from the cluster X-ray
centroid. These include three extended radio galaxies (two with the
narrow-angle tail morphology and one a double-lobed source), one marginally
resolved object (a possible ``dying'' radio galaxy, as discussed by
\citealt{2014A&A...561A..52V}), and 12 unresolved sources.  We then subtracted
the CLEAN components associated with these compact sources (for a total flux
density of 75 mJy) from the $uv$\/ data and used the resulting dataset to
obtain images of only the diffuse radio halo at multiple resolutions using the
multi-scale CLEAN.  Images restored with a $22''$ circular beam before and
after the removal of the sources unrelated to the halo are shown in
\autoref{fig:radio}. The image with sources removed (panel {\em a}) has an rms
noise level of 22~$\mu$Jy~beam$^{-1}$. The halo flux density, measured within
the $1\sigma$ isocontour, is $20.2$ mJy with an error of 7.2\%, computed
following \cite{2013ApJ...777..141C}, i.e., including the flux calibration
uncertainty (3\%), the noise in the integration area, and the error due to the
subtraction of the discrete radio sources in the halo region. Our flux
measurement is in agreement with the flux density of $19.8\pm1.4$ used by
\cite{2013ApJ...777..141C} to calculate the radio halo luminosity at 1.4 GHz
and measured on an image obtained from the AF349 observations. A slightly lower
flux of $16.7\pm0.6$ mJy is measured by \cite{2014A&A...561A..52V} by masking
the radio galaxies (rather than subtracting them as we did).

Our high-quality image of the radio halo reveals a prominent edge and
significant brightness structure on small angular scales. We will compare
these fine features with our X-ray data in \autoref{sec:radio_features} and
\autoref{sec:radio_origin}. As we will see, the dataset still exhibits
some interferometric artifacts with the scale and amplitude that are significant
for us; we will address this in \autoref{sec:radio_modelling}.

\section{Discussion}

\subsection{Bow Shock}

\begin{figure*}[!ht]
        \centering
        \leavevmode
        \includegraphics[width=85mm,keepaspectratio,bb=50 160 535 640,clip]
                {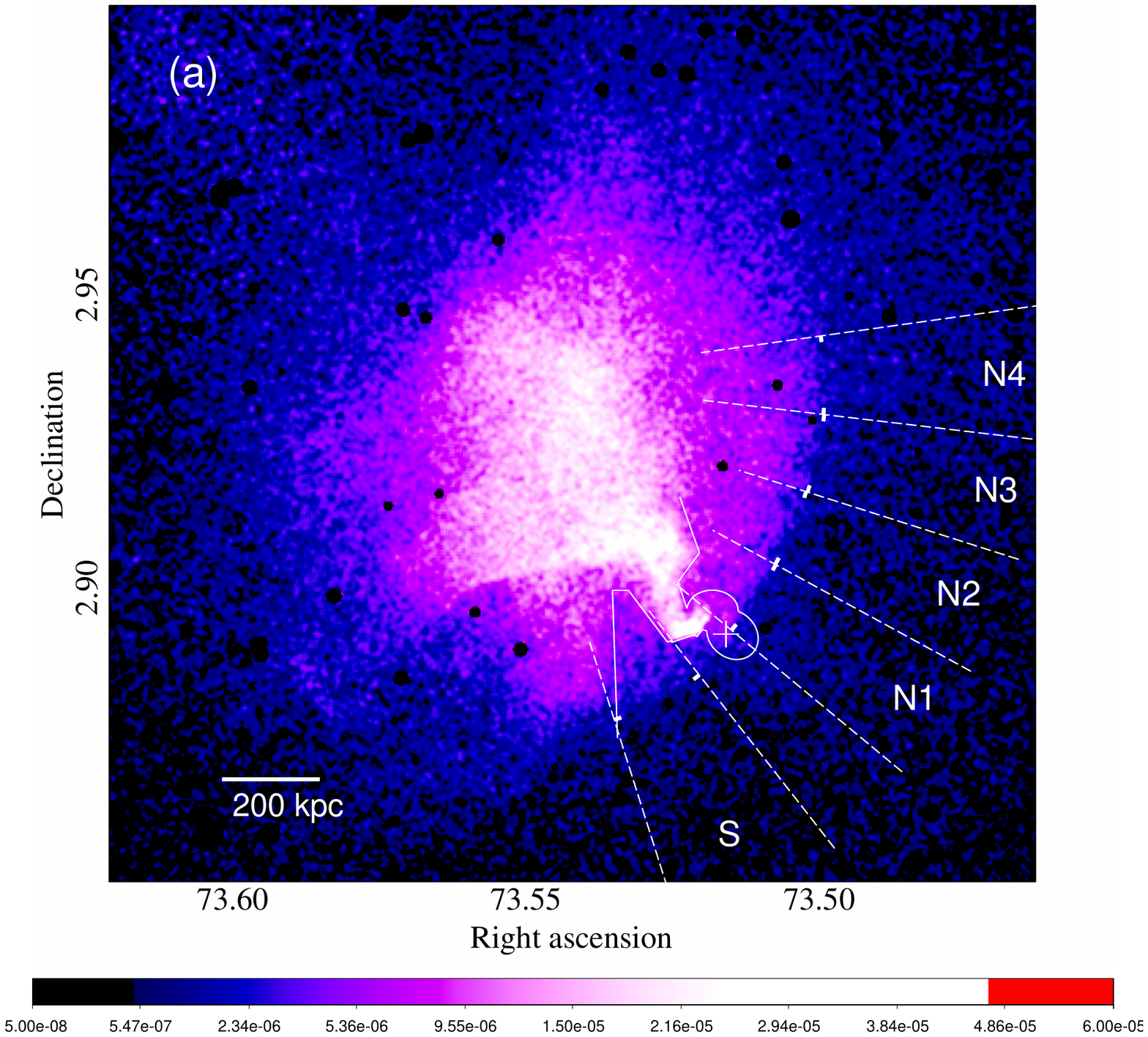}
        \hfil
        \includegraphics[width=85mm,keepaspectratio,bb=0 7 252 250,clip]
                {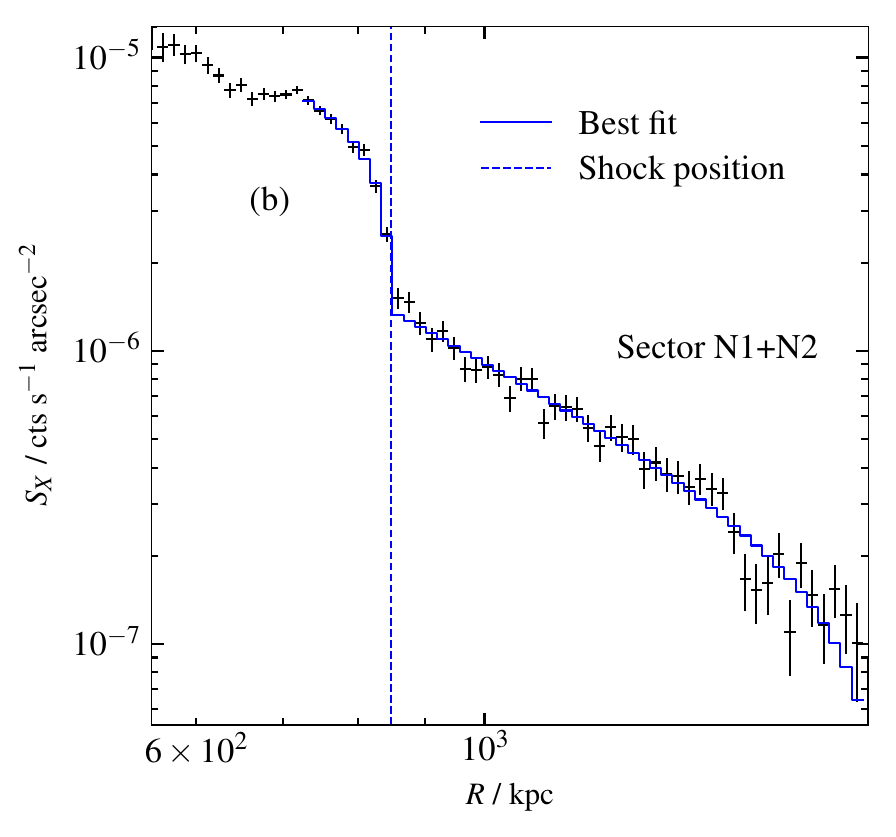}

        \caption{
          ({\em a}) 0.8--4~keV \chandra\ image, smoothed by 2\as\ gaussian
          (holes are masked point sources). The white dashed lines mark the
          sectors used for X-ray shock profile fitting, with tick marks
          indicating the best-fit shock position in each sector. The white
          cross indicates the position of the BCG next to the cold front. The
          white solid outline indicates masked regions for the X-ray brightness
          profile and spectral extraction, covering the cold front close to
          the shock surface.
          ({\em b}) X-ray brightness profile in the combined sector N1+N2 and
          best-fit model.
          200~kpc is 1\am.\\}
        \label{fig:xray}
\end{figure*}

A classic bow shock would exhibit the highest Mach number $M$\/ (and the
highest gas density and temperature jumps) at the ``nose,'' and a decreasing
$M$\/ as the angle from the axis of symmetry of the shock increases and the
shock becomes oblique. For the electron--proton equilibration test that we want
to perform, we need as high a Mach number as possible to maximize the
difference in electron temperature between the two possibilities, and thus
want to study as narrow a sector at the ``nose'' as possible. The deep A520
X-ray observation provides sufficient statistics to analyze the bow shock in
several sectors, divided based on the brightness contrast across the front
(\autoref{fig:xray}{\em a}). We exclude a narrow segment of the front between
sectors S and N1 immediately in front of the bright cool structure because of
the small-scale irregularities, possibly caused by the dark matter mass peak
located there (see Figure 6 in \citetalias{2016ApJ...833...99W}), that would be
difficult to model. A small region that includes those structures is also
excluded from sector N1 as shown in \autoref{fig:xray}{\em a}.

In each sector, we fit the 0.8--4.0 keV surface brightness profile with a
density model that consists of two power-law radial profiles (with different
slopes) on either side of the shock and an abrupt jump at the shock, whose
position is a free parameter. This 3D model is projected onto the sky under
the assumption that the curvature along the line of sight (l.o.s.)  is the
same as that of the brightness edge in the plane of the sky (which is further
discussed in \autoref{sec:geom_unc}) and compared to the brightness profiles
extracted in the respective sector. For these observations, we can use the
0.8--4.0 keV count rate as a direct proxy for the l.o.s.-integrated $n_e^2$,
because the combination of the spectral model parameters ($N_H$, abundance,
gas temperatures) and the \chandra\ response in this energy band conspire to
make the dependence of the X-ray flux on temperature negligible ($<1$\% for
the interesting range of temperatures, based on examining how the predicted flux responds to varying the plasma temperature of the model in XSPEC). The Mach number, $M$, of the shock
front relative to the upstream flow is derived from the density jump $x$\/
using the Rankine-Hugoniot jump conditions \citep{1959flme.book.....L}:
\begin{equation}
M = \left( \frac{2x}{(\gamma+1)-(\gamma-1)x} \right)^{1/2},
\end{equation}
where we use the adiabatic index $\gamma = 5/3$ for monatomic gas. For
typical, low Mach numbers found in clusters, $x$\/ is far from the
asymptotic value for strong shocks (4 for $\gamma = 5/3$) and thus allows an
accurate determination of $M$. Fit results for the sectors are shown in
\autoref{tab:xrayfit}. The highest Mach numbers,
$M=2.5_{-0.4}^{+0.5}$
and
$M=2.4_{-0.3}^{+0.6}$,
are seen in sectors N1 and N2 at the ``nose'' of the shock, respectively, and
decrease to $M<2$ on either side, where the shock becomes oblique. The shock
``nose'' direction is in agreement with the NE--SW merger axis. The ``nose''
sectors have higher values compared to
$M=2.1_{-0.3}^{+0.4}$
reported in \citetalias{2005ApJ...627..733M} because the latter included the
adjacent sectors with lower density jumps.

Although the $M$\/ decline toward the wings of the front is expected, it has
only been reported previously for the main shock of the Bullet cluster
\citepalias{2007PhR...443....1M}, because such a study requires good
statistics. Care should therefore be taken when using a wide sector to analyze
bow shocks, as the peak density jump will probably be underestimated.

\begin{table*}[t]
\caption{Details of X-Ray Modeling of Shock Sectors}
\begin{center}
\begin{tabular}{lcccccccccc}
\hline\noalign{\smallskip}
\hline\noalign{\smallskip}
Sector & $\rho$ Jump & $M$ & Inner slope & Outer slope & $\chi^2/\nu$  \\
\noalign{\smallskip}
\hline\noalign{\smallskip}
S & $2.0^{+0.2}_{-0.3}$ & $1.7^{+0.2}_{-0.2}$ & $-0.8^{+0.4}_{-0.6}$ & $-1.8^{+0.3}_{-0.2}$ & $59.5/43$  \\
N1 & $2.7_{-0.3}^{+0.3}$ & $2.5_{-0.4}^{+0.5}$ & $-0.3_{-0.7}^{+0.7}$ & $-1.7_{-0.2}^{+0.3}$ & $35.6/27$ \\
N2 & $2.6^{+0.4}_{-0.2}$ & $2.4^{+0.6}_{-0.3}$ & $-0.2_{-0.6}^{+0.7}$ & $-1.7_{-0.3}^{+0.2}$ & $34.3/27$ \\
N3 & $2.2^{+0.3}_{-0.2}$ & $1.9^{+0.3}_{-0.2}$ & $-1.1^{+0.6}_{-0.8}$ & $-1.7^{+0.2}_{-0.2}$ & $44.2/27$  \\
N4 & $1.9^{+0.2}_{-0.2}$ & $1.6^{+0.2}_{-0.1}$ & $-0.4^{+0.3}_{-0.3}$ & $-1.5^{+0.1}_{-0.2}$ & $68.5/39$ \\
N1+N2 & $2.7^{+0.2}_{-0.3}$ & $2.4^{+0.4}_{-0.2}$ & $-0.5^{+0.8}_{-0.7}$ & $-1.6^{+0.1}_{-0.2}$ & $24.7/30$ \\
\noalign{\smallskip}
\hline{\smallskip}
\end{tabular}
\end{center}
Notes. -- Column (1): shock sector as shown in \autoref{fig:xray}{\em a}; column
(2): density jump at shock; column (3): shock Mach number; column (4): density
profile inner power law index; column (5): density profile outer power law
index; column (6): chi-square and degree of freedom. Errors are 90\% with all
other parameters free.
\label{tab:xrayfit}
\end{table*}

\subsection{Electron--Ion Equilibration Timescale}
\label{sec:equilibration}

In the collisional plasma picture, a shock front with relatively low Mach
numbers --- such as those occurring in cluster mergers --- would heat protons
and heavier ions dissipatively, while electrons, whose thermal velocity is
much higher than the velocity of such shocks, are compressed adiabatically to
a temperature lower than that of ions. Protons and electrons subsequently
equilibrate on a Coulomb collision timescale (e.g.,
\citealt{1962pfig.book.....S}; \citealt{1966egct.book.....Z}):
\begin{equation}
\tau_{\text{ep}} = 2 \times 10^8 \text{ yr } \left( \frac{n_e}{10^{-3} \text{cm}^{-3}} \right)^{-1} \left( \frac{T_e}{10^8 \text{ K}} \right)^{3/2}.
\label{eq:tep_spitzer}
\end{equation}
We can measure the electron temperature directly by modeling the X-ray
spectrum, but cannot measure the proton temperature. However, the {\em
equilibrium}\/ post-shock temperature $\overline{T_2}$\/ (the one that protons
and electrons achieve asymptotically) can be derived from the pre-shock
temperature and the compression factor $x$\/ (or, equivalently, the Mach
number) using the shock jump conditions:
\begin{equation}
\frac{\overline{T_2}}{T_1} = \frac{(\gamma+1)x-\gamma+1}{x(\gamma+1)-x^2(\gamma-1)}.
\end{equation}
Indices 1 and 2 correspond to pre-shock and post-shock quantities,
respectively. The time dependence of the electron temperature $T_e$\/
increasing asymptotically from the adiabatic value to the equilibrium value
under Coulomb collisions was given by, e.g., \cite{1997ApJ...491..459F},
\cite{2009ApJ...707.1141W}, and \cite{2016arXiv160607433S}. As the local
$T_e$\/ increases, the post-shock gas flows away from the shock front, and the
$T_e$\/ time dependence gets encoded in the spatial temperature profile, which
can be measured by \chandra. The closing piece of the experimental setup is
the velocity of the post-shock gas relative to the front, which is given by
the shock mass conservation condition:
\begin{equation}
v_2 =M c_{s1} / x,
\label{eq:v2}
\end{equation}
where $c_{s1}$\/ is the pre-shock sounds speed, determined from the X-ray
measured pre-shock temperature (electrons and ions in the pre-shock region can
reasonably be assumed in equilibrium).

The Mach number should be sufficiently high to distinguish between shock
heating and adiabatic compression. For $M=2.4$, there is a measurable
difference between the two, but they become practically indistinguishable for
$M \lesssim 2$. In sectors N1 and N2, the shock is strongest, and their Mach
numbers are statistically the same (\autoref{tab:xrayfit}), thus we will
combine them (see the N1+N2 entry) and use the combined profile for the above
test. We will use the other sectors, which should be insensitive to the
possible temperature non-equilibrium, for consistency checks.

We construct two model $T_e$\/ profiles for each sector, one for adiabatic
compression at the shock and subsequent Coulomb equilibration, and the other
for instant equilibration, and compare them with the observed temperature
profile. The electron density is derived from the emission measure using the
normalization of the {\sc APEC} model in XSPEC,
$\int n_e n_H dV =10^{14} \times \text{norm}\times 4 \pi [D_A (1+z)]^2$.
In sector N1+N2, we find
$n_{e1}=(4.04\pm0.15)\times 10^{-4}$~cm$^{-3}$
immediately in front of the shock, and
$n_{e2}=(1.07\pm0.11)\times 10^{-3}$~cm$^{-3}$
behind the shock (the latter includes a 10\% uncertainty of the density jump).
Reasonable deviation of the shock surface from spherical does not affect our
results -- this is discussed in \autoref{sec:geom_unc}.

\begin{figure}[!ht]
        \includegraphics[width=85mm,keepaspectratio,bb=173 256 438 535,clip]
                {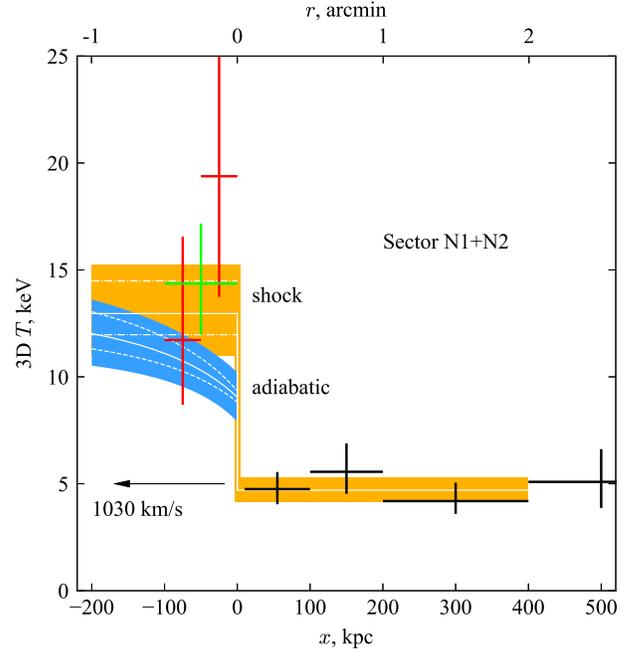}

        \caption{ Deprojected post-shock temperatures compared with model
        profiles in sectors N1 and N2 combined (\autoref{fig:xray}{\em a}).
        The yellow band is the instant-equilibration model, while the blue band is
        adiabatic compression followed by Coulomb equilibration. The band
        width indicates 1-$\sigma$ error bounds. In the pre-shock region, this
        equals the error in the pre-shock temperature measurement, while in
        the post-shock region this further includes the density jump parameter
        uncertainty, which has a smaller effect.
        The white dashed and dotted--dashed lines bounding the post-shock model
        profiles indicate the effect of geometric uncertainty ($\pm$10\%
        change in the l.o.s. extent).
        The $x$-axis denotes distance from the shock position.
        Different colors of post-shock crosses correspond to bins of different
        widths (red is 50~kpc and green is 100~kpc). $x$-error bars denote the
        radii in which temperature was measured. $y$-errors are 1$\sigma$. }
        \label{fig:t_eq_deproj}
\end{figure}

The pre-shock temperature profile for N1+N2 is consistent with being constant
out to at least 800~kpc from the shock (\autoref{fig:t_eq_deproj}). We thus
decided to use the best-fit temperature in a radial bin between 10 and 400~kpc
from the shock (where we excluded the immediate vicinity of the shock to avoid
any irregularities of the front),
$T=4.70^{+0.82}_{-0.72}\text{ keV}$,
as the pre-shock value. This temperature, the best-fit compression factor, and
\autoref{eq:v2} give the post-shock gas velocity of
$1030_{-80}^{+90}$~km~s$^{-1}$
relative to the shock front. During the collision equilibration timescale
$\tau_{\text{ep}}$~$\simeq$~0.2~Gyr,
the post-shock gas travels $\sim$200~kpc or 65\as, which is well resolved by
\chandra.

\begin{figure*}[!htp]
        \centering
        \leavevmode
        \includegraphics[width=175mm,keepaspectratio,clip]
                {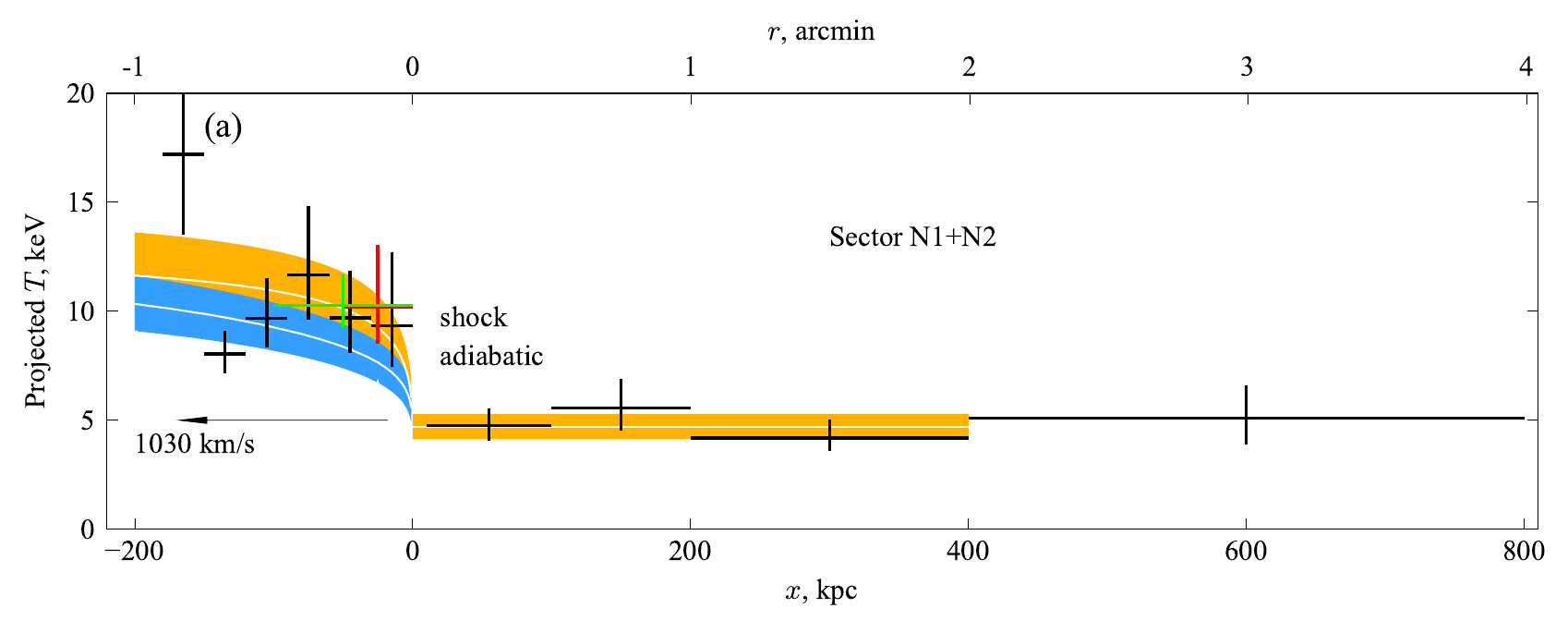}
        \centering
        \leavevmode
        \includegraphics[width=175mm,keepaspectratio,clip]
                {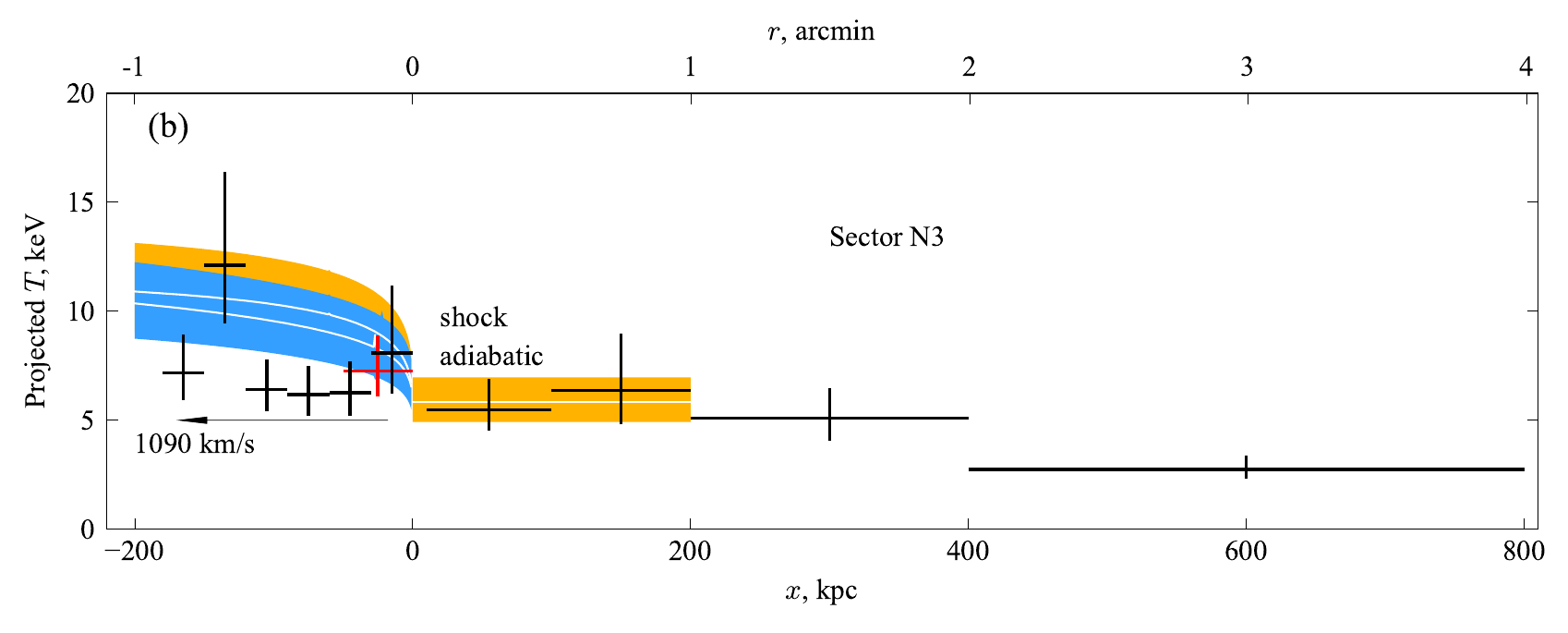}
        \centering
        \leavevmode
        \includegraphics[width=175mm,keepaspectratio,clip]
                {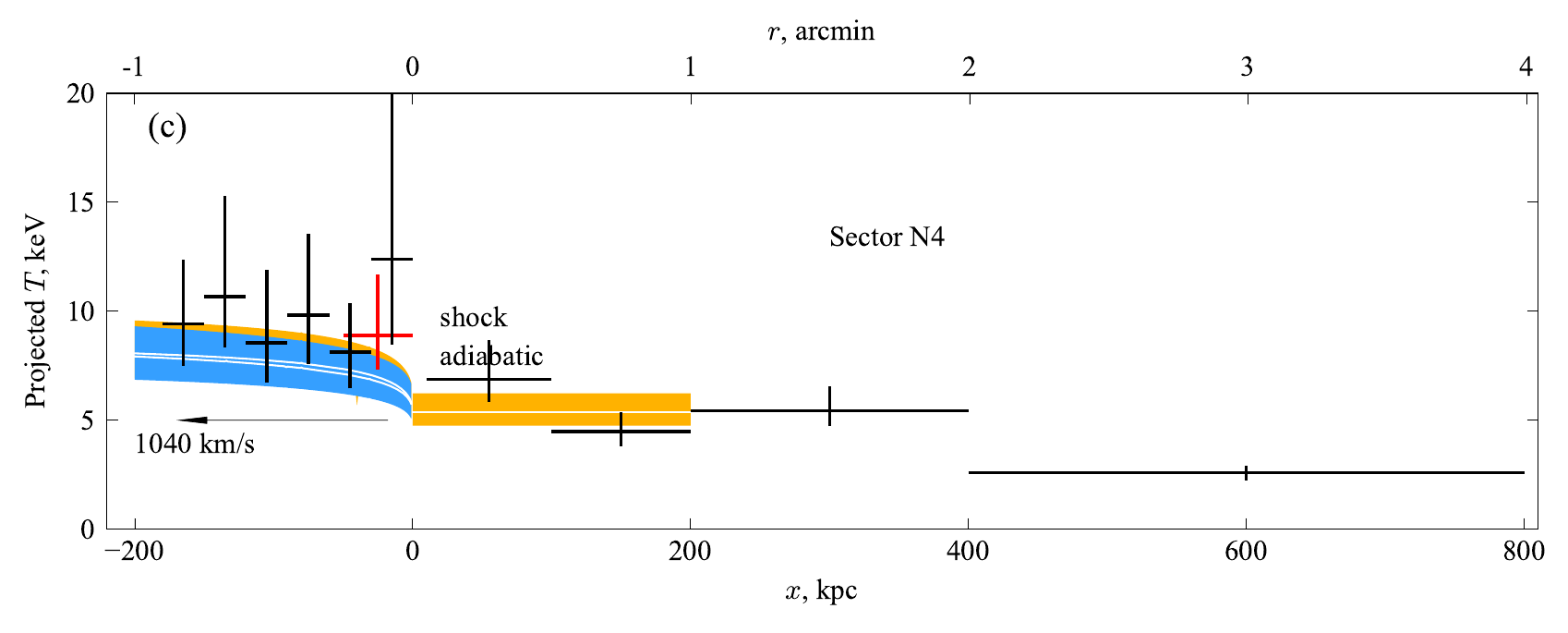}

        \caption{ Projected temperatures compared with model profiles for
        segments of the shock surface in various sectors labeled in
        \autoref{fig:xray}{\em a}.
        ({\em a}) Projected spectroscopic-like temperature in sector N1+N2 (using $n^2
        T^{-3/4}$ weighting; following
        \citealt{2004MNRAS.354...10M}).
        ({\em b}) Projected temperature in sector N3.
        ({\em c}) Projected temperature in sector N4.
        The $x$-axis denotes distance from the shock position. Different
        colors of post-shock crosses correspond to bins of different widths.
        $x$-error bars denote the radii in which temperature was measured.
        $y$-errors are 1$\sigma$. The bands are 1$\sigma$ error bounds.}
        \label{fig:t_eq_proj}
\end{figure*}

We will compare the deprojected and projected temperature profiles for N1+N2
with the instant-equilibration and adiabatic compression models in
\autoref{fig:t_eq_deproj} and \autoref{fig:t_eq_proj}{\em a}. To construct the
adiabatic compression model, we calculate the time dependence of the local
post-shock $T_e$\/ using the measured shock parameters following
\cite{2016arXiv160607433S}. For the instant-equilibration model, we assume the
electron temperature jumping to $\overline{T_2}$ right at the shock. These
models with their uncertainties, which include statistical uncertainties of
the pre-shock temperature and the density jump, are shown in
\autoref{fig:t_eq_deproj}. The 3D temperature model profile is projected onto
the sky using the best-fit density model and the spectroscopic-like
temperature weighting
$w=n^2 T^{-3/4}$,
following \cite{2004MNRAS.354...10M}. The projected model profiles are shown
in \autoref{fig:t_eq_proj} for sector N1+N2 as well as N3 and N4.

We also deproject the {\em measured}\/ temperatures in bins of several sizes
(30, 50, 100~kpc) immediately after the shock by estimating the
contributions of the outer 3D shells into the spectrum from the respective
post-shock region and adding the properly normalized spectral component in the
XSPEC fit to represent the projected gas. Using projected profile and using deprojected
profile are equivalent --- the difference between the two models is greater
for the 3D profiles, but so is the uncertainty of the deprojected measured
temperature.

For the N1+N2 sector, in the 50~kpc bin behind the shock, we measure the
projected temperature of
$T=10.2_{-2.4}^{+5.3}\text{ keV}$,
while the deprojected temperature using the best-fit 3D density model is
$T=19.4_{-8.4}\text{ keV}$
(unconstrained on the high side because of \chandra's poor sensitivity to such
high temperatures). The first post-shock bin is the most useful, because it
has the greatest model difference. For illustration purposes, we also obtained
a deprojected temperature for the second 50~kpc shell, with the first 50~kpc
post-shock shell fixed at the model instant-equilibration temperature, in
effect deprojecting the instant-equilibration model (this is done to
regularize the deprojection procedure, as the errors of the neighboring bins
are anti-correlated). The procedure was repeated for three narrower 30~kpc
post-shock bins, and for one wider 100~kpc bin.

The deprojected $T_e$ in the first 30, 50, and 100~kpc post-shock bins
are all above the adiabatic model and consistent with the
instant-equilibration model. (Of course, these measurements are not
statistically independent.) The adiabatic model is below the measured
deprojected value at 95\% significance in a single parameter test for the
50~kpc and 100~kpc bins, and around 90\% for the 30~kpc bin. Similarly, the
projected spectroscopic-like temperatures are higher than the adiabatic
compression model at 95\% significance for the 50~kpc and 100~kpc bins.

In the 30 and 50~kpc bins further from the shock, the temperatures remain
consistent with the instant-equilibration model. In
\autoref{fig:t_eq_proj}{\em a}, we show the 30~kpc bins for the projected
profile up to about 200~kpc behind the shock to give a broader overview;
however, cool core fragments and unrelated cluster structure can be seen
within 200~kpc behind the shock (\autoref{fig:xray}{\em a}), which can affect
the projected temperatures and create the apparent large scatter.

For a consistency check, we also obtained the projected temperature profiles
in sectors N3 ($M=1.9^{+0.3}_{-0.2}$) and N4 ($M=1.6^{+0.2}_{-0.1}$), where
the Mach numbers are insufficiently high to distinguish the two models
(\autoref{fig:t_eq_proj}{\em b,c}). In these sectors, the pre-shock
temperature shows a slow decrease with radius, so we derived the best-fit
pre-shock temperatures in a narrower 10--200~kpc bin. They are consistent with
the pre-shock temperature for N1+N2. In both sectors, the temperature increase
immediately behind the shock is consistent with both models. In N3, the
presence of a cool blob of gas causes measurements from about 50~kpc behind
the shock to fall down; this blob has been seen in the temperature map (Figure
2 in \citetalias{2016ApJ...833...99W}). In N4, measurements appear
systematically above the models (although not significantly). This may be
caused by our underestimating of the immediate pre-shock temperature, as the
deviation at the first pre-shock bin suggests. We conclude that sectors N3 and
N4 behave consistently with the expectation for the lower Mach numbers
observed in these sectors.

\subsubsection{Geometrical Systematic Uncertainty}
\label{sec:geom_unc}

In the above experiment, we relied on the assumption that the shock surface
has the same curvature along the line of sight as in the plane of the sky.
This is a reasonable assumption for this merger with a relatively clear
geometry, for which the apparent shock direction is generally well aligned
with the merger axis evident from both the X-ray and lensing maps, and the
shock front center of curvature in the image approximately coincides with the
large-scale cluster centroid. Nevertheless, we should determine how the
uncertainty of this assumption affects the results. If the surface has a
different curvature along the l.o.s., we would derive the incorrect density
jump and Mach number. The deprojected post-shock temperature would also be
affected, but because of the relatively high brightness contrast at the shock
(i.e., a relatively low projected contribution), this is a secondary effect.

To evaluate the effect, we varied the radius of curvature of the shock surface
along the l.o.s., while keeping the pre-shock gas model unchanged
(spherically symmetric). For simplicity we used a spheroid geometry for the
shock surface, keeping its axes in the plane of the sky to be the same as the shock
radius $r_{\text{jump}}$, while linearly stretching its l.o.s.\ axis. Note that
with this geometry, the extent $l$ of the shock surface in the l.o.s.\
direction scales with the l.o.s. radius of curvature $R$ not linearly but as
$l \propto \sqrt{R}$.

For a 20\% change in $R$, the best-fit density jump changed by 5\% --- the
difference coming from the change in post-shock density, while the pre-shock
density stays the same. This is smaller in magnitude than the $\sim$10\%
fitting error on this parameter, so for a moderate amount of shock-surface
variation, the geometry does not significantly affect our results (see
\autoref{fig:t_eq_deproj}). For this uncertainty to become dominant, the shock
surface should be very asymmetric, e.g. a factor 1.7 different $R$\/
corresponds to a 15\% change to the density jump.

\subsubsection{Comparison with Other Shocks}
\label{sec:other_shocks}

While such a degree of asymmetry seems unlikely for the relatively symmetric
merger in A520, there is no way of knowing this for sure for each individual
shock. One way to assess the probability of the shock front asymmetries and
how well the true Mach number is recovered from the X-ray density profiles is
to study shocks in cosmological simulations. On the observational side,
measurements for a sample of relatively strong ($M \gtrsim 2.5$) shocks is
needed for a robust conclusion on the electron--proton equilibration timescale.
Our A520 result adds a data point to two other previously published
measurements --- the Bullet cluster with $M\approx 3$
(\citetalias{2006ESASP.604..723M}, \citetalias{2007PhR...443....1M}) and the
stronger of the two shocks in A2146, one with $M=2.3$
(\citetalias{2012MNRAS.423..236R}).  The Bullet cluster showed, at a similar
95\% confidence, a similar preference for fast electron--proton equilibration.
The A2146 shock showed preference for the Coulomb equilibration at a similar
$\sim$2$\sigma$ significance (considering, as we do, only the temperature bin
immediately after the shock) -- although the instant-equilibration and Coulomb
models themselves were only $1\sigma$ apart due to a low $M$\/ and a large
uncertainty for the pre-shock temperature. The physics of the intracluster
plasma in different clusters should be similar, so we should get the same
answer from all of the experiments. The mild contradiction between the Bullet and
A520 on one side and A2146 on the other may be a reflection of the above
geometrical uncertainty. We do note that the bow shock in A2146 used in
\citetalias{2012MNRAS.423..236R} exhibits a flat shape at its ``nose,'' with
the shock center of curvature far from the cluster centroid (see Figure 8 in
\citetalias{2012MNRAS.423..236R}), which diminishes our confidence in the
above l.o.s./image plane symmetry assumption. It is interesting that their
second, weaker shock exhibits a $T_e$ jump that is {\em higher}\/ than the
prediction of both models (at a similar significance), which may be further
illustration of the geometric uncertainty. An additional apparent difference
between our (along with \citetalias{2006ESASP.604..723M} and
\citealt{2016arXiv160607433S}) and \citetalias{2012MNRAS.423..236R} analyses
is the three times longer Coulomb equilibration timescale used in
\citetalias{2012MNRAS.423..236R} (cf.\ their Equation 2 and our
\autoref{eq:tep_spitzer}), although this would not reconcile the results.

\cite{2016arXiv160607433S} performed a similar test on an $M=2.5$ shock front
at the position of the western radio relic in A3667. Their derived post-shock
electron temperature goes below even the adiabatic model, which would appear
to indicate a problem with this method. However, that shock is located 2 Mpc
away from the cluster center,
where the cluster emission is very faint and a projection of any unrelated
X-ray structure on the line of sight may have a significant effect. For
example, if a faint, cool group were projected onto the post-shock region,
it would result in both an overestimate of the gas density jump and an
underestimate of the temperature jump --- effect of the right sign to
explain their result. For A520, as well as the Bullet and A2146, projection of
unrelated objects is much less of a problem because the shocks are
located in much brighter cluster regions.

Thus, our conclusion is score 2:1 in favor of quick electron--proton equilibration in the
intracluster plasma, but more strong shocks need to be studied to reduce the
systematic uncertainties.

\subsection{Radio Halo Features}
\label{sec:radio_features}

\begin{figure*}[!ht]
        \centering
        \leavevmode
        \includegraphics[width=85mm,keepaspectratio,bb=50 160 535 640,clip]
                {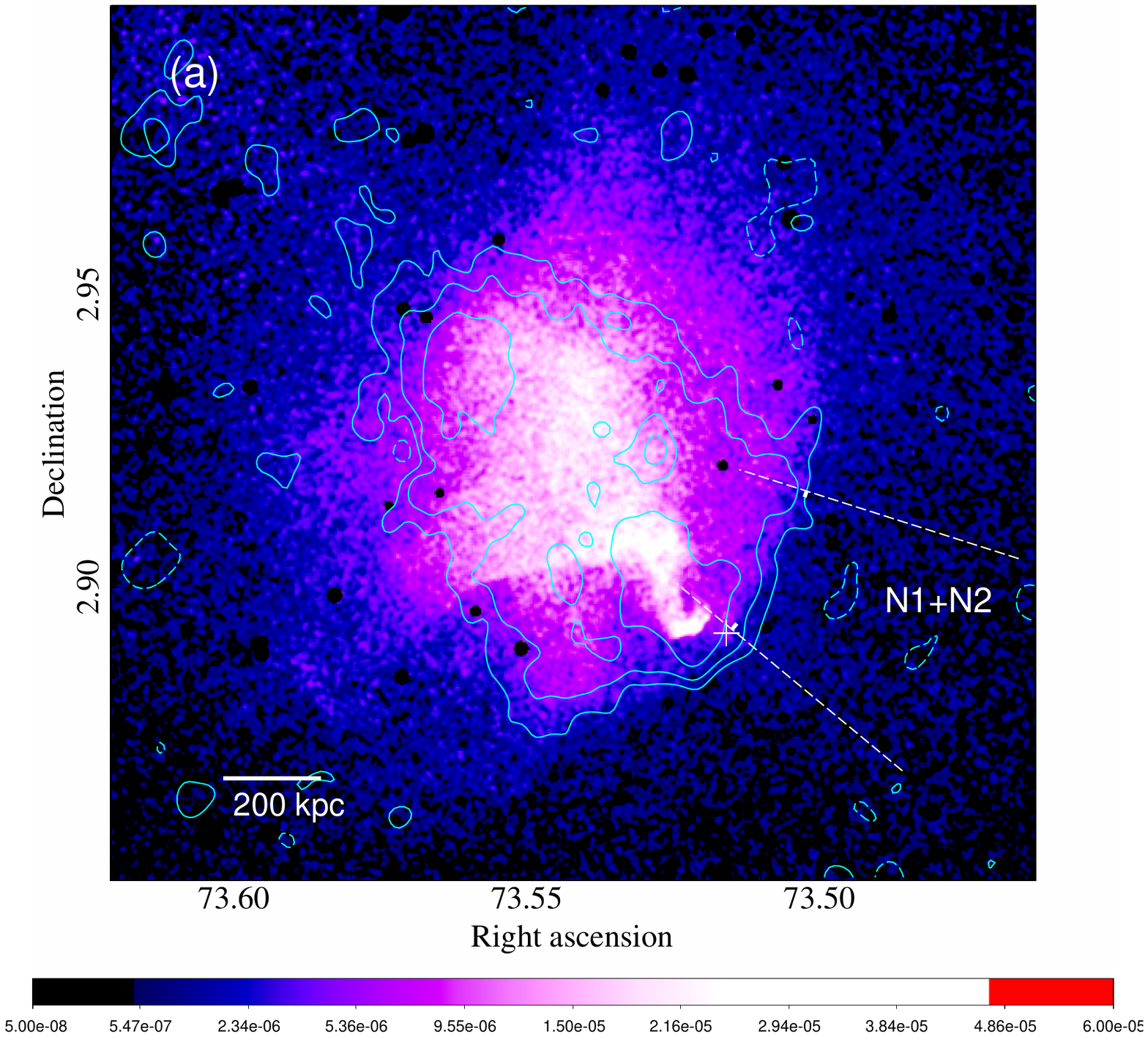}
        \hfil
        \includegraphics[width=85mm,keepaspectratio,bb=0 8 250 250,clip]
                        {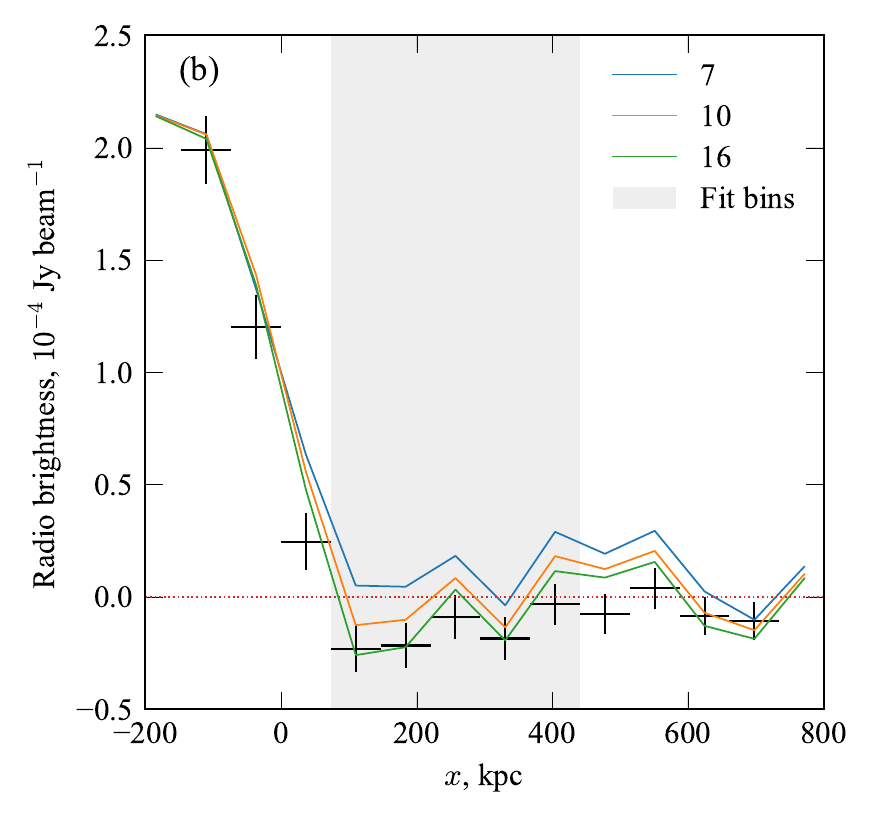}

        \caption{({\em a}) 0.8--4~keV \chandra\ image, same as
          \autoref{fig:xray}{\em a}, showing the combined sector N1+N2. Radio
          contours are the same as in \autoref{fig:radio}{\em a}.
          ({\em b}) Radio brightness profile in the combined sector N1+N2 (crosses: extracted radial profile; solid lines: profiles extracted from simulated images with additional pre-shock emission injected, for different values of the emissivity jump; grey band: the 5 bins used to measure $\Delta \chi^2$, see \autoref{sec:radio_modelling}). The radial bins correspond to the FWHM beam size.
          $x$ is relative to the X-ray best-fit shock position. 200~kpc is
          1\am.\\}
        \label{fig:radio2}
\end{figure*}

There are interesting coincidences between the radio halo and X-ray features
in A520 (\autoref{fig:radio2}{\em a}). There are bright radio spots at the positions
of the cool ``foot'' and ``knee'' that we discussed in
\citetalias{2016ApJ...833...99W}. The radio emission here may be related to a
radio {\em minihalo}\/ that had inhabited the cool core before its disruption,
which gave rise to these cool X-ray clumps, as minihalos are observed in
almost all massive cool cores \citep{2017ApJ...841...71G}. The radio
enhancements there may also be caused by reacceleration of relativistic
particles by local turbulence in the wake of the disrupted cool core (see
\citealt{2014IJMPD..2330007B} for review of possible acceleration mechanisms
in clusters). Another prominent, broad brightness peak in the NE half of the
radio halo is located at one of the hottest regions of the cluster, but it
does not have any obvious corresponding X-ray brightness structures. This can
be the site of vigorous merger-induced turbulence, which would produce
relativistic electrons via reacceleration. Future spatially resolved X-ray
calorimeters, such as {\em XARM}, will be able to detect this turbulent
region.

\subsection{Origin of the Radio Edge}
\label{sec:radio_origin}

As discussed in \citetalias{2005ApJ...627..733M}, the X-ray bow shock in A520
traces a sharp edge of the radio halo, and we see it clearly in
\autoref{fig:radio2}. Mechanisms of producing ultra-relativistic electrons
responsible for the post-shock radio synchrotron emission include first-order
Fermi acceleration, which can use thermal electrons as its seeds or
re-accelerate ``fossil'' relativistic electrons (e.g.,
\citealt{1987PhR...154....1B}) that existed prior to shock passage but whose
radio brightness is below the detection limit. Another possible mechanism is
adiabatic compression of such fossil electrons and the compression of the
magnetic field (since cosmic rays and magnetic fields are frozen into the
thermal gas that is being compressed by the shock). Both the adiabatic
compression and the reacceleration should be present, but the reacceleration
boost for aged, steep-spectrum electrons depends on the the Lorentz factor
$\gamma_{\rm max}$ of the fossil electrons (see
\citetalias{2005ApJ...627..733M} for discussion of the resulting spectrum and
normalization), so either effect may dominate. In either of these two scenarios,
the fossil relativistic electrons in the pre-shock region should produce radio
emission at a certain low brightness level that can be related to that of the
post-shock emission. As derived in \citetalias{2005ApJ...627..733M} in the
compression-only scenario, for a power-law fossil electron energy spectrum
with index $\delta$ (defined as $dN/d\gamma\propto
\gamma^{-\delta}$), a gas density jump by factor $x$\/ at the shock, and
certain assumptions about the tangled magnetic field, the radio emissivity per
unit volume would change as
\begin{equation}
I_{\nu} \propto x^{\frac{2}{3}\delta +1}.
\label{eq:inu}
\end{equation}
If both compression and significant reacceleration are present, for a fixed
{\em observed}\/ post-shock radio brightness, we would expect a lower level of
pre-shock radio emission, and in the case of the Fermi acceleration directly
from the thermal pool, the pre-shock radio emission would be lower still by
many orders of magnitude. With our new, higher-sensitivity radio map, we can
try to test these possibilities by extracting a radio surface brightness profile
across the shock.

\subsubsection{Modeling Radio Emissivity Profile}
\label{sec:radio_modelling}

In the same sector N1+N2 where we obtained the highest Mach number bins, we
extracted a radio profile binned to the beam size and aligned with the
best-fit shock position. It is shown in \autoref{fig:radio2}{\em b}; the radio brightness
drops sharply at the position of the X-ray shock and is not detected in the pre-shock region.
To evaluate measurement errors for the profile, we
generated Gaussian noise images with the observed rms noise of
22~$\mu$Jy~beam$^{-1}$ after smoothing by the beam size, and extracted radial
profiles from 1000 smoothed noise images. An elongated ``hole'' in the radio
image in the N1+N2 sector about 200~kpc in front of the shock
(dashed radio contour in \autoref{fig:radio2}{\em a}) is most likely an interferometric
artifact. Since we want to place an {\it upper} limit on the pre-shock radio
emission, to be conservative we masked this negative deviation.

We will now compare this radio brightness profile and, in particular, the
non-detection in the pre-shock region, with the expectation for an adiabatic
compression model for the origin of the radio edge.
To model the radio image, we created a spherical model of the radio
emissivity in the relevant region of space, projected it on the sky, and
convolved it with the {\em VLA}\/ beam.

For lack of information on the distribution of cosmic rays and magnetic fields in A520,
our model makes two assumptions. First, the density of cosmic-ray electrons
is assumed proportional to that of thermal ICM. If we consider the various possible sources for
fossil electrons --- merger shock acceleration and subsequent vigorous mixing,
disrupted and mixed radio galaxies, turbulent acceleration, and ``secondary''
electrons from cosmic-ray proton collisions (see \citealt{2014IJMPD..2330007B}
for a review) --- this seems a reasonable assumption. We note that for our purpose,
this is a conservative assumption compared to the alternative of a flat cosmic-ray
density profile. Second, we assume
the magnetic field strength changes across the cluster as $B \propto n^{0.5}$\/ (where $n$\/
is gas density), which is the best fit derived for Coma \citep{2010A&A...513A..30B},
also a merging cluster.
Then the synchrotron emissivity (emission per unit volume) $P \propto nB^2
\propto n^2$, so the pre-shock radio emissivity has essentially the same dependence
on the gas density as the X-ray. We therefore use the radial profile derived from
the X-ray and only let the normalization change to model the radio profile.

The post-shock region is fit very well with the projection of a 3D model with an abrupt emissivity drop, convolved with the beam. We assume constant emissivity vs.\ radius in this region,
because we are most interested in the bin immediately next to the shock surface
and because it is not clear how the radio brightness should change further downstream (the image does not have enough leverage to fit this slope because of the unrelated radio structures inside the cluster halo).
We did check that the post-shock radio profile does not favor, for example, a thin 3D shell that in
projection would show a peak at the shock position and a decline toward smaller radii.
The jump in radio emissivity at the X-ray shock location is the only free parameter. The model is
truncated at 1.5~Mpc from the X-ray centroid (which is 1.15~Mpc from the shock surface);
because of the model's steep decline, this does not matter much.

Because an interferometer can lose signal on large angular scales, we must
be careful when deriving an upper limit in the low-surface-brightness areas. The unknown zero
level of the image limits our ability to constrain the pre-shock emission,
but we should be able to constrain models with steep changes on linear scales that
are well within the nominal {\em uv}\/ coverage limit, which is $\sim 3$ Mpc for this dataset (\autoref{sec:radio}).
However, there may be subtle artifacts on all scales, and for example,
the apparent systematic negative values in the radio profile in the pre-shock region
are a cause for concern. With this in mind, rather than simply fitting the projected radio
emission model to the profile, we tried to account for the possible artifacts
to a first approximation by convolving the brightness model with the
actual {\em uv}\/ coverage and the beam and reconstructing the image. Technically, we followed
\cite{2014ApJ...795...73G} and ``injected'' or added our brightness model for the pre-shock
emission into the pre-shock region of the data using the AIPS task UVSUB.
We then extracted a radial profile in the same sector of the new image, thought
of it as a model, and compared it with the actual profile using the $\chi^2$ statistics.
(Because the same statistical noise is present in both the real and the ``model'' images,
for the $\chi^2$ calculation we used the errors for only one of the profiles). The
injected brightness model was calculated by keeping the post-shock emission at the same best-fit
level, while varying the jump amplitude and thus the normalization of the pre-shock profile
(only the pre-shock region of the model emission was injected). We compared the data
and model profiles in the 400 kpc pre-shock radial interval \autoref{fig:radio2}{\em b} to avoid being
affected by the accuracy of our model assumptions while being interested only in
the shock jump.

This exercise revealed that the negative deviations in the pre-shock region are
indeed an artifact --- the difference between the image with and without
the injection there was {\em less}\/ than the injected emission, which means that
the interferometer does redistribute the flux from this region into other radial
bins (see, e.g., the positive bump around $x\approx 500$ kpc). Theoretically, it should
be possible to account for this effect and constrain the absolute brightness in the
pre-shock region, but it would require creating an accurate spatial model of the radio brightness
for the entire cluster, which is beyond the scope of this study. (It may be more
efficient instead to obtain a dataset with better {\em uv}\/ coverage that would not
require such modeling.)

Nevertheless, we can evaluate the sensitivity of this radio image to the pre-shock
emission under the assumption that the true pre-shock emission in the data is zero.
Then, an injected model that corresponds to the emissivity jump by a factor of 10 (see \autoref{fig:radio2}{\em b})
is rejected at a $3\sigma$ statistical significance, while a jump by a factor of 16
is rejected at $2\sigma$. If we ignored the interferometric artifact and simply
convolved the brightness model with the beam without accounting for the
{\em uv}\/ coverage, we would have excluded at a $3\sigma$ level a jump by factor 22.

\subsubsection{Comparison with Adiabatic Compression}

Let us now compare this with the emissivity jump expected in the adiabatic compression model.
Compression should preserve the shape of the electron energy spectrum, while
shifting it in frequency and changing its normalization. For a power-law
electron spectrum, the radio synchrotron spectrum
$I_{\nu}\propto \nu^{-\alpha}$
is related to the electron spectrum via
$\alpha = (\delta -1)/2$.
For the radio spectral index, we use
$\alpha = 1.25 \pm 0.11$
($1\sigma$ errors) from the first post-shock bin in Figure 6 of
\cite{2014A&A...561A..52V}, which corresponds to
$\delta = 3.5 \pm 0.2$.
Formally, for the observed gas density jump of $x=2.7$ in sector N1+N2
(\autoref{tab:xrayfit}),
\autoref{eq:inu} gives the expected radio emissivity jump of
$27 \pm 4$
for the adiabatic compression scenario.
For comparison with the data, we need to include projection effects, because the
post-shock radial brightness profile includes regions along the line of sight
that are away from the shock ``nose,'' with a lower density jump.
The gas density jump azimuthal dependence in the plane of the sky in the sectors in which it was
measured, can be interpolated well by
$x = x_{\mathrm{nose}} (\cos \theta)^{1/2}$,
where $\theta$ is the angle from the ``nose'' of the shock. Then, we assumed
rotational symmetry about the ``nose'' and calculated
the shock-surface-area-weighted radio jump (given by \autoref{eq:inu}) in the
first post-shock bin in the radio brightness profile.
This gives a value of 16 for the average radio emissivity jump, which can be directly
compared with the limits derived above for a model that did not
include this azimuthal dependence for simplicity.

If we approximately include radiative cooling of the post-shock
relativistic electrons, we expect a lower, more
easily detectable emissivity jump at the shock. The radio spectrum should
steepen within $\sim$100~kpc downstream of the shock
(\citetalias{2005ApJ...627..733M}). The beam size of the \vla\ data used to
calculate spectral index images by \citeauthor{2014A&A...561A..52V} is 130~kpc
(see their Figure 4), so such a spectral change in the immediate post-shock
region is not resolved, and the spectral index immediately at the shock would
be flatter. An unresolved mixture should have a volume-averaged slope of
$\bar{\alpha}\approx \alpha+1/2$
\citep{1964ocr..book.....G}, so for $\bar{\alpha} \simeq 1.25$ the value at
the shock could be $\alpha \simeq 0.75$, for which the radio emissivity jump
in the compression scenario would be a factor of 9 (including the
projection of oblique shock contributions), compared to 16 obtained above
without cooling. The true value may be somewhere in this
interval, depending on the processes in the post-shock plasma. Note that in
the above calculations, we do not consider radiative cooling of the pre-shock electrons, in effect assuming either that something balances that cooling
or that the pre-shock electrons are continuously generated by some process (e.g.,
cosmic-ray proton collisions with thermal protons). If cooling is balanced
pre-shock, it may be balanced post-shock as well, so the above cooling
correction for the spectral index would not be necessary. A high-resolution
map of the post-shock spectral index may shed light on the relevant physical processes here.

Comparing these estimates to the limits above, we see that the statistical
sensitivity of the radio data would allow us to exclude such jumps at
$>$3$\sigma$ confidence. However, because of the unfortunate interferometric
artifact, the exclusion significance is lower, only $\sim$2$\sigma$, and it
depends on the assumption about the zero level in the image. Nevertheless,
this demonstrates that ruling out the compression model is within reach
with a dataset with similar sensitivity but better {\em uv}\/ coverage.

Note that our estimates above used the assumption that the
electrons have a power-law energy spectrum. If this is not the case (e.g.,
both pre-shock and post-shock spectra may have a cutoff at some frequencies
because of radiative cooling), the adiabatic model can still be
constrained, but it requires measurements at several frequencies.
As noted in M05, using our notation, a single electron
 emits most of its synchrotron radiation at a frequency that scales with the
 compression factor as
\begin{equation}
\nu_{\mathrm{peak}} \propto B\gamma^2 \propto x^{4/3}.
\end{equation}
For $x=2.7$, the post-shock electron emitting at
1.4~GHz would have emitted at 370~MHz before the shock passage (or at 560~MHz for
$x=2.0$). So pre-shock observations at those lower frequencies, combined with the post-shock 1.4 GHz
brightness, would be least dependent on the assumed shape of the electron spectrum.
Alternatively, pre-shock measurements at 1.4 GHz would need to be combined with
higher-frequency data for the post-shock region. And, ultimately, measuring the
spectrum of the post-shock emission in the relevant range above and below 1.4 GHz
and verifying that it is a power law (or detecting a curvature) would
provide the most robust constraint.

\section{Summary}

We analyzed a deep \chandra\ exposure of A520 to study its prominent bow
shock, one of only a handful of merger shocks with simple and unambiguous
geometry and a relatively high Mach number. At the ``nose'' of the shock, we
find $M=2.4_{-0.3}^{+0.4}$. This is higher than in the previous study based on
a shorter exposure (\citetalias{2005ApJ...627..733M}), because we were able to
use a narrower sector at the ``nose'' of the shock. As expected, the Mach
number declines (toward 1.6--1.7) away from the ``nose,'' where the shock
front becomes oblique.

The relatively high Mach number of the central segment of the front allowed us
to perform a test of the electron--proton equilibration timescale, similar to
the earlier tests for the Bullet cluster (\citetalias{2006ESASP.604..723M},
\citetalias{2007PhR...443....1M}) and A2146
(\citetalias{2012MNRAS.423..236R}). We fit the shock X-ray brightness profile
using a gas density model with a jump, used the density jump to evaluate the
post-shock gasdynamic temperature, and compared it to the measured post-shock
electron temperature. The electron temperature immediately behind the shock is
higher than expected from a simple picture where electrons are compressed
adiabatically by the shock and then equilibrate with protons on a Coulomb
collisional timescale. This indicates a faster equilibration rate, pointing to
the prevalence of other particle interactions in hot magnetized plasma.
Although the confidence level is only 95\% (this includes the statistical
error on temperature, ACIS background uncertainty, and sky background effect),
it is similar to the finding for the Bullet cluster
(\citetalias{2006ESASP.604..723M}, \citetalias{2007PhR...443....1M}). Although
the A2146 result (\citetalias{2012MNRAS.423..236R}) was inconclusive (mostly
because its Mach number is lower and the amplitude of the effect is smaller),
it did prefer adiabatic compression over fast equilibration. The scatter
between these results most likely reflects the geometric uncertainty inherent
in this test
--- the curvature of the shock front in the sky plane is used to model its
curvature along the line of sight. This scatter can be averaged out by
studying a sample of shocks, and our result provides a third entry for
such a sample. Unfortunately, bow shocks that are as clear-cut as Bullet
or A520 are rare, so expanding the sample may require going to higher
redshifts with more sensitive instruments. This is worth the effort,
because cluster shocks provide one of the most direct methods of determining this
important timescale for any astrophysical plasmas.

We also present a new combined analysis of the archival 1.4 GHz radio \vla\
data on the cluster giant radio halo, previously analyzed separately in
\cite{2001A&A...376..803G} and \cite{2014A&A...561A..52V}. In addition to
providing lower statistical noise, the datasets complement each other's
interferometric coverage, which improves fidelity of the reconstructed image.
The radio image reveals several interesting features, such as the bright spot
that coincides with the disrupted cool core, possibly related to a former
minihalo. Another bright spot may point to a region of high turbulence, a
possible target for future X-ray calorimetric measurements.

A520 is one of the growing number of clusters where both a giant radio
halo and an X-ray shock front are observed \citep{2012mgm..conf..397M}.
As in most of them, there is a prominent sharp edge of the radio halo that
coincides with the X-ray shock front. Some clusters have X-ray shocks with
counterparts both in the form of the halo edges and radio relics
\citep{2015MNRAS.449.1486S}.
Studying these colocated features may shed light on the physical
processes responsible for the generation and acceleration of the
radio-emitting electrons. For example, in our A520 dataset, the radio
emission in the pre-shock region is undetected at a very low
brightness level, which has not been probed for any other shocks.
If the jump of the radio emission at the shock were caused by simple
adiabatic compression of relativistic electrons in the pre-shock plasma (e.g., remaining from past shocks or produced throughout the cluster by
cosmic-ray proton interactions), we should see the radio emission
beyond the edge (\citetalias{2005ApJ...627..733M}). We came close to being able
to rule this model out (and thus demonstrate the
existence of particle acceleration or reacceleration at shocks) based on statistical
sensitivity of the radio data. However, an interferometric artifact
in the region of interest dominates the uncertainty. Our analysis shows that
this interesting test for the cluster radio halos is within reach, but probably
requires an observation with a better interferometric coverage and at
lower frequencies, e.g., with {\em GMRT}\/ or {\em LOFAR}.

\acknowledgements
We thank the anonymous referee for valuable criticisms that led us to
more accurate conclusions.
Q.H.S.W. was supported by \chandra\ grants GO3-14144Z, GO5-16147Z, and AR5-16013X.
S.G. acknowledges the support by the National Aeronautics and Space
Administration, through Chandra Award Numbers AR5-16013X and G05-16136X. Basic
research in radio astronomy at the Naval Research Laboratory is supported by
6.1 Base funding.

\bibliography{cluster2}

\end{document}